\documentclass[10pt,letterpaper]{article}
\usepackage[top=0.85in,left=1.5in,footskip=0.75in]{geometry}

\usepackage{changepage}

\usepackage[utf8]{inputenc}

\usepackage{textcomp,marvosym}

\usepackage{fixltx2e}

\usepackage{amsmath,amssymb}

\usepackage{cite}

\usepackage{nameref,hyperref}

\usepackage{microtype}
\DisableLigatures[f]{encoding = *, family = * }

\usepackage{rotating}

\raggedright
\usepackage[aboveskip=1pt,labelfont=bf,labelsep=period,justification=raggedright,singlelinecheck=off]{caption}

\makeatletter
\renewcommand{\@biblabel}[1]{\quad#1.}
\makeatother

\date{}

\usepackage{lastpage,fancyhdr,graphicx}
\usepackage{epstopdf}
\begin{document}
\vspace*{0.35in}

\begin{flushleft}
{\Large
\textbf\newline{Measuring Verifiability in Online Information}
}
\newline
\\
Reed H. Harder\textsuperscript{1},
Alfredo J. Velasco\textsuperscript{2},
Michael S. Evans\textsuperscript{3,4,*},
Daniel N. Rockmore\textsuperscript{3,5,6}
\\
\bigskip
\bf{1} Thayer School of Engineering, Dartmouth College, Hanover, New Hampshire, United States of America
\\
\bf{2} Department of Electrical and Computer Engineering, Duke University, Durham, North Carolina, United States of America
\\
\bf{3} Neukom Institute for Computational Science, Dartmouth College, Hanover, New Hampshire, United States of America
\\
\bf{4} Department of Film and Media Studies, Dartmouth College, Hanover, New Hampshire, United States of America
\\
\bf{5} Department of Mathematics, Dartmouth College, Hanover, New Hampshire, United States of America
\\
\bf{6} Department of Computer Science, Dartmouth College, Hanover, New Hampshire, United States of America
\\
\bigskip


%
%
%
%
%
%

* Corresponding Author: michael.evans@dartmouth.edu

\end{flushleft}
\section*{Abstract}
The verifiability of online information is important, but difficult to assess systematically. We examine verifiability in the case of Wikipedia, one of the world's largest and most consulted online information sources. We extend prior work about quality of Wikipedia articles, knowledge production, and sources to consider the quality of Wikipedia references. We propose systems to calculate technical accuracy and practical accessibility of sources. We calculate article verifiability scores for a sample of 5,000 articles, and compare differently weighted models to illustrate effects of emphasizing particular elements of verifiability over others. We find that, while the quality of references in the overall sample is reasonably high, verifiability varies significantly by article, particularly when emphasizing the use of standard digital identifiers and taking into account the practical availability of referenced sources. We discuss the implications of these findings for measuring verifiability in online information more generally.



\section*{Introduction}

With the rise of widely-available networked communication, individual and social life increasingly relies on online sources of information~\cite{Castells:1996ys}. Potential medical patients self-diagnose by searching online medical databases. Teachers and students use online reference material to provide education in mathematics, science, history, and art. Scientific researchers build on published work and online datasets to create new knowledge and advance our understanding of the world. 

The quality of online information is therefore an important public and scholarly concern. This concern is reflected in scholarly literature examining information quality from a variety of system and user perspectives. Studies examine, for example, how consumers seek health information on the Internet~\cite{Cline:2001rc}, whether online drug information is accurate~\cite{Clauson:2008rw}, whether Internet news sites reinforce existing political beliefs~\cite{Garrett:2009eq}, and how scientific misinformation persists online~\cite{Kata:2010fr}.

This study addresses this general concern about online information by examining one specific quality of online information: verifiability. As we use the term here, ``verifiability'' means the extent to which information can be checked for reliability, truth content, or accuracy. We borrow the term from Wikipedia, which is a free online encyclopedia and one of the most frequently consulted websites in the world~\cite{Alexa:2015sf}. As a collaboratively written and edited encyclopedia with more than 24 million contributors worldwide, Wikipedia is also a complex sociotechnical system and ``knowledge instrument'' that relies on a variety of highly structured policies and voluntary enforcement to preserve the integrity of knowledge being conveyed~\cite{Wikipedia:2015qf,Niederer:2010qq}.

\subsection*{Technical and practical verifiability}

Though Wikipedia is much larger and extensive than many online information sources, it provides an illustrative example of the challenges to quality that many online information sources face. At the heart of Wikipedia's collaborative processes are the ``core content policies'' of ``verifiability,'' ``no original research'' and ``neutral point of view.'' The most basic of these is verifiability. In Wikipedia, verifiability is the foundation of reliable knowledge. According to Wikipedia policy documents, ``[a]ll material in Wikipedia mainspace, including everything in articles, lists and captions, must be verifiable.'' For policy purposes, verifiability ``means that people reading and editing the encyclopedia can check that the information comes from a reliable source''~\cite{Wikipedia:2015ij}.

As with many online information sources, the most obvious challenge to verifiability in Wikipedia is a lack of citations and references. Without any reference material, it is difficult to verify whether information is true, accurate, and reliable. Thousands of words of Wikipedia policy documentation address the maintenance of verifiability through the correct use of references and citations to reliable sources. Current instructions focus on how to identify when citations are missing, how to provide those citations in such cases, and how to determine whether provided citations meet the Wikipedia standards for reliability~\cite{Wikipedia:2015eu}.

But it is important to note that simply providing citations and references does not automatically guarantee verifiability. Whether or not provided references and citations are accessible is less often considered as a challenge to verifiability. But it is just as important as providing the reference or citation in the first place. There are many ways that an online information source might provide citations and references and still be difficult to verify. These possible challenges fall into two analytical categories: ``technical verifiability'' and ``practical verifiability.'' 

``Technical verifiability'' is the extent to which a reference provides supporting information that permits automated technical validation of the existence of the referenced material, based on existing technical standards or conventions. For example, books can be located with International Standard Book Number (ISBN) or Google Books ID, and journal articles can be located with a Digital Object Identifier (DOI). A missing ISBN or DOI certainly makes it more difficult to locate a book or article. But a provided ISBN or DOI could also be invalid or even entirely fictional, rendering the reference useless for verifying the information it supports. Thus, a Wikipedia article, all of whose book and journal references were invalid, would not be ``technically verifiable'' under this definition. Note that technical verifiability thus does not speak to the usefulness or relevance of the referenced material, just its existence. In particular, if all the ISBNs and DOIs corresponded to existing materials, but were mistakenly attached to the article, the article would still be perfectly `technically verifiable,' although upon deeper inspection, clearly failing by some other measure.

``Practical verifiability,'' by contrast, is the extent to which referenced material is accessible to someone encountering the reference. For example, if a DOI is present but refers to a paywalled journal article, then the information it supports is practically unverifiable to someone without the additional means to access the supporting journal article. Similarly, if an ISBN is present but refers to a book that only has one extant copy in a library thousands of miles away, then the information it supports is practically unverifiable to someone without the additional means to access the supporting book.


In what follows we examine technical verifiability and practical verifiability in popular Wikipedia articles. We examine two research questions about verifiability using data from Wikipedia. Our first research question (hereafter ``RQ1'') asks, are existing citations verifiable (for multiple types of basic identifiers)? For our second research question we construct some simple verifiability metrics and then ask the question (hereafter ``RQ2''), do different versions of a verifiability metric produce different rankings? We address each of these questions using a sample of the top 5,000 English-language Wikipedia articles and the citations/references that those articles contain. We address RQ1 by quantifying and measuring various elements of verifiability, such as the presence or absence of identifiers. We address RQ2 by creating simple weighted sum metrics obtained by applying different weights to different elements of verifiability, observing the changes in ranking among these highly consulted Wikipedia articles, and considering how different verifiability scores might reflect differences in how articles are constructed.


\subsection*{Information quality in Wikipedia}

Our analysis of Wikipedia extends a significant amount of related work on the quality or qualities of Wikipedia. This related work falls into three broad categories: quality of articles in terms of accuracy and error; quality of the editing process in terms of error correction and updating; and quality of sources in terms of reliability and diversity.

Several small-scale studies assess the quality of individual articles. For example, a 2005 investigation by \textit{Nature} found that 21 selected articles from Wikipedia were comparable in quality to their \textit{Encyclopedia Britannica} counterparts~\cite{Giles:2005sf}. However, a study of 9 Wikipedia articles on specific historical topics found that the sampled articles lacked the accuracy of \textit{Encyclopedia Britannica} and the detail of \textit{American National Biography Online}~\cite{Rector:2008fh}. And a recent study of Wikipedia articles on the 10 most costly medical conditions found that Wikipedia articles on 9 of the 10 conditions showed ``statistically significant discordance'' from peer-reviewed literature~\cite{Hasty:2014jo}.

Related studies focus on the quality of the knowledge production process in Wikipedia. For example, a 2008 study inserted incorrect information (``fibs'') into a sample of ``well-tended'' Wikipedia articles and found that somewhere between one third and one half of these ``fibs'' were corrected within 48 hours~\cite{Magnus:2008lq}. An investigation of editorial disputes over the Wikipedia verifiability policy found that the structural organization of Wikipedia editing largely prevents oligarchic takeover by a few influential editors (or a ``cabal'')~\cite{Konieczny:2009mz}. On a broader scale, a meta-analysis of existing studies concluded that the ``epistemic virtues'' of Wikipedia knowledge production generally outweigh the drawbacks~\cite{Fallis:2008bs}.

A third group of studies examines the quality of sources (e.g., references and citations) in Wikipedia articles. At the smaller scale, a study of a random sample of 50 country history articles from Wikipedia found that articles tended to refer to online sources, and disproportionately relied on news media and government websites~\cite{Luyt:2010dk}.  In a larger scale project, a study of reference editing activity in a sample of 137,104 articles found that more mature articles are more likely to have more extensive references~\cite{Chen:2012cl}. And at a much larger scale, a study of 11 million citations in Wikipedia found that US sources are most common, that Google, media companies such as the New York Times, and databases such as IMDb and Census.gov dominate citations, and that primary sources are among the most persistent (and therefore ``most valued'') in Wikipedia articles~\cite{Ford:2013qq}.

We extend this related work, and its concern with Wikipedia information quality, by providing a method of assessing the quality of \textit{references}. We evaluate whether journal and book citations that are actually used in Wikipedia are valid, accessible, and available to a wide range of users and editors. We also demonstrate how this method can be used to construct a verifiability score for any (or every) Wikipedia article. This method can be further extended to any online information source that provides references to support verifiability.

So, while this paper draws primarily on data from Wikipedia, the analysis also reflects and informs current concerns about standardization and access in information systems well beyond Wikipedia. Our analysis of technical verifiability reflects and informs ongoing concerns about the evaluation and reliability of data quality at scale \cite{boyd:2012th}. Our analysis of practical verifiability reflects and informs ongoing concerns about openness, access, and digital inequality \cite{Suber:2012fj}. By creating and testing a verifiability metric that accounts for many different verifiability failure modes, that can be applied at different scales, and that accommodates different weights on different versions of verifiability for assessment purposes, we advance the methodology of verifiability measurement in online information sources more generally.

%

\section*{Analysis}



\subsection*{Data}

Wikipedia makes regular data dumps of its content available for download. We extracted 22,843,288 citations from the 3,437,650 citation-containing articles in the English Wikipedia data dump made on July 7th, 2014. Wikipedia keeps a data dump of the number of visits each article receives per hour~\cite{Wikipedia:2015db}. We aggregated the page views for each hour of the entire year and took the top 5,000 most viewed (as of July 2014) whose titles were found among the 3,437,650 citation-containing articles in the English Wikipedia. 

The article sampling strategy reflected two analytical objectives. First, we wanted the sample to contain actively viewed articles rather than unmaintained or idle articles that were unlikely to motivate maintenance activity. Second, we wanted the sample to contain a range of articles in terms of official quality, rather than only focusing on the best (``featured'') articles in Wikipedia. Some top articles are featured articles of enduring interest, but many are low-quality (``stub'') articles that cover subjects of fleeting popularity.

Wikipedia does not strictly enforce a particular format for citations~\cite{Wikipedia:2015eu}. However, several commonly used markup methods account for the majority of references in articles. Inline citations, corresponding to specific lines of text in the article, are usually formed using the ``{\textless}ref{\textgreater}'' tag in Wikipedia markup, which contains additional information about the source, often including reference type (book, journal, etc.), link if available, and other document identifiers. Citations can also appear that are not anchored to any particular piece of text: we refer to these as ``free'' citations. Free citations usually are marked with one of several common citation templates. 

Our citation extraction pulled both inline citations and free citations from articles. Citations were categorized by citation type, either book or journal. Book citations were checked for the presence of ISBNs or other identifying information. Journal citations were checked for DOIs and other numerical identifications.


\subsection*{RQ1: Are existing citations verifiable?}

\begin{figure}
\centering
    \includegraphics[width=\columnwidth]{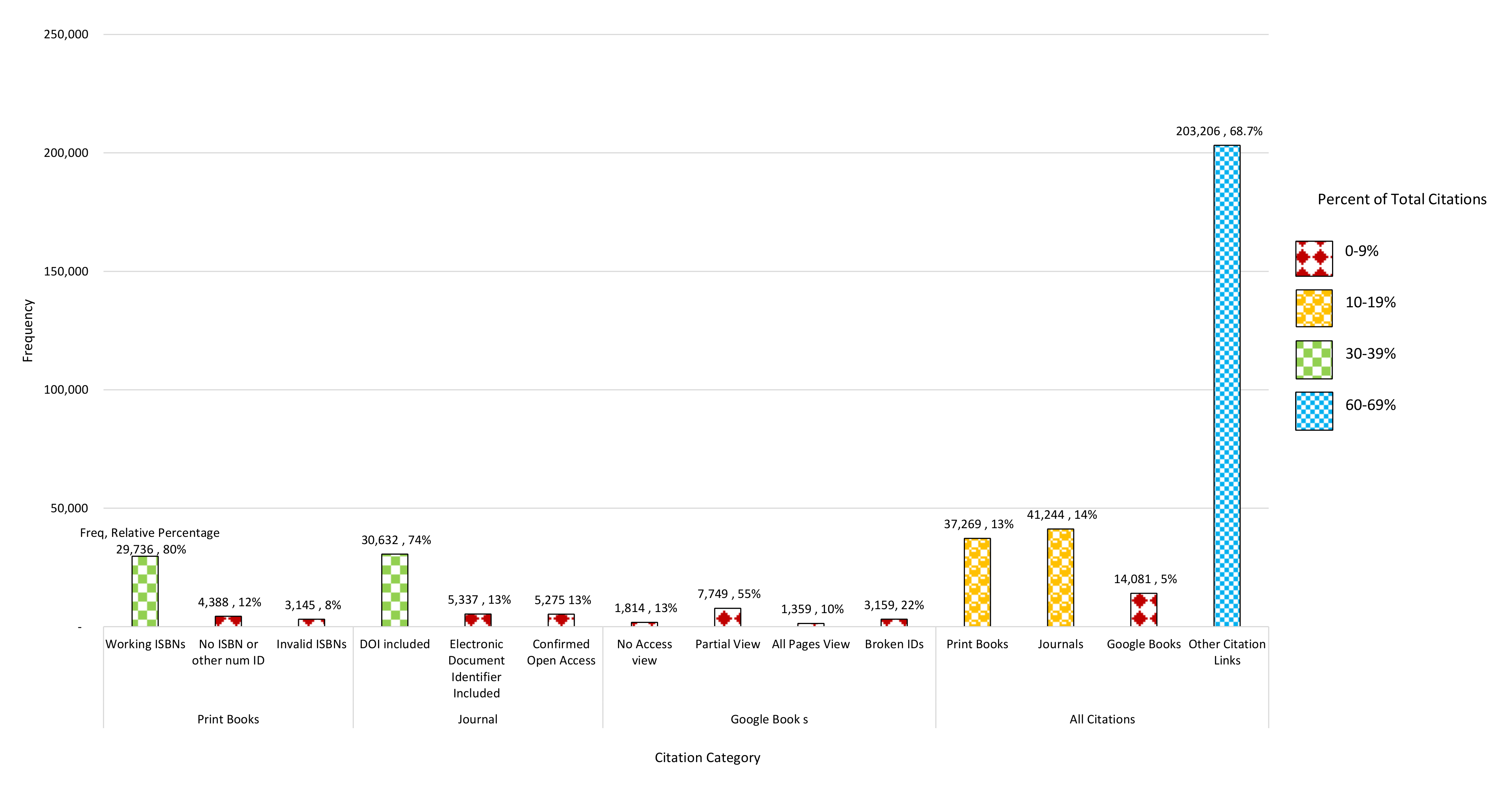}
  \caption{\bf Citation frequency breakdown by category for the 5,000 most visited English articles.}~\label{fig:figure1}
\end{figure}

Fig.~\ref{fig:figure1} displays the citation frequency breakdown by category for the 5,000 most visited English articles. Our analysis focuses on book and journal citations that are potentially covered by standardized numerical identifier systems. For numerical identifiers we focused on ISBNs, Google Books IDs, and DOIs. We checked to see whether any of these identifiers were present. If they were present, we checked for validity. In the case of Google Books IDs and DOIs, we also checked the extent to which the linked resource was freely available for viewing by an ordinary user of Wikpedia.

\paragraph{Technical Verifiability}

Fig.~\ref{fig:figure1} includes measures of technical verifiability for books and articles. ISBN numbers are the standard publishing industry identifier for books. ISBN numbers can be checked numerically for validity using check-digit algorithms for either their 10 or 13 digit versions~\cite{Hahn:2015uq}. ISBNs found with Wikipedia citations in the `book' reference type specified in the Wikipedia markup were tested according to these algorithms. Out of 37,269 book citations, 29,736 book citations  (79.8\%) had valid ISBNs, while 3,145 (8.4\%) of book citations had invalid ISBNs, and 4,388 book citations (11.8\%) contained no ISBN information. 

An alternative standardized book identifier is Google Books ID. Google Books IDs were extracted from references containing Google Books links. This process did not rely on the `book' reference type being indicated in Wikipedia markup, as this markup is inconsistent across references. Links were tested for validity using bulk submissions to a Google developer API designed for Google Books~\cite{Google:2015qv}. Out of 14,081 Google Books-containing citations, 3,159 (22.4\%) contained invalid Google Books IDs.

Adding the presence of valid Google Books IDs as a marker of technical verifiability even in the absence of a valid ISBN, we get a slight improvement in the overall technical verifiability of book citations: 31,578 (84.7\%) contain valid identifiers, 3,218 (8.6\%) lack valid identifiers, and 2,473 (6.6\%) contain invalid identifiers. Adding in consideration of Google Books links in other citations (not explicitly labeled ``book''), we see similar proportions: 34,231 (84.7\%) out of 40,381 contain valid identifiers, 3,218 (8.0\%) lack valid identifiers, and 2,932 (7.3\%) contain invalid identifiers.

Journal article citations were slightly more difficult to test for validity in bulk form. Instead, presence or absence of a Digital Object Identifier (DOI) was noted for any reference tagged as `journal', `study', `dissertation', `paper', `document', or similar. Out of 41,244 of these citations, only 5,337 (12.9\%) contained neither a DOI or a link to a known open access journal.

\paragraph{Practical Verifiability}

Fig.~\ref{fig:figure1} also includes measures of practical verifiability for articles and books. Verifying the open access nature of a journal citation beyond the simple presence or absence of a digital identifier is often difficult. Only a few journals are exclusively open access, and journal reference pages often have idiosyncratic layouts, making bulk web scraping for open-access confirmation challenging. Journal citations linking to `arXiv' and `PubMed Central (PMC)' were taken to be open access, while all others were marked unconfirmed. 5,275 of the journal citations out of 41,244 (12.8\%) belonged to this confirmed open access category, while 30,632 or 74.3\% contained some digital identifier but were not confirmed to be open.

Google's API allowed us to classify the accessibility of the linked Google Books into three categories: fully viewable, with all pages accessible; partially viewable, with a sample available; or not viewable at all. Out of the 10,922 working Google Books links, most (7,749, or 71.0\%) are partially viewable with samples, while 1,359 (12.4\%) are fully viewable and 1,814 (16.6\%) are not viewable at all.

\subsection*{RQ2: Do different versions of a verifiability metric produce different rankings?}

In order to formulate and test different metrics for the verifiability of Wikipedia articles, we took proportions from the technical and practical verifiability measures calculated above, and took a weighted sum to produce an aggregate score for each page. For measures of technical validity, we looked at the proportion of valid ISBNs, and the proportion of functional Google Books identifiers. For measures of practical verifiability, we looked at the proportion of journals verifiably open access (in arXiv and PMC), the proportion of linked Google Books with fully open access, and the proportion of linked Google Books with partial access. We also considered presence or absence of numerical identifiers: the proportion of journals with a DOI, and the proportion of book citations with some sort of numerical identification (either from Google or an ISBN). 

Using these measures, we constructed 4 different models of aggregate scoring, each weighting different proportions more or less heavily. Our baseline model (referred to as Model 1) weighted the technical and practical aspects of verifiability equally (with partial Google Books access conferring half the weight of a full Google Books access). So, for example, in our baseline model the article ``arbitration'' received a score of 2.07, ``Bugatti'' received a score of 3, and ``Nero'' received a score of 2.27. Table \ref{tab:table1} displays the score breakdown under Model 1.

\begin{table}[ht]
\caption{
{\bf Examples of score calculation, Baseline model (Model 1).}}
\begin{tabular}{r c c c c}
\textbf{Score component} & \textbf{(Weight)} & \textbf{Arbitration} & \textbf{Bugatti} & \textbf{Nero}\\
ISBNs valid & 1 & 1 & 1 & 1 \\ \hline
Google Books links valid & 1 & 1 & 1 & 1 \\ \hline
journals with DOI & 0 & 0 & 0 & 0 \\ \hline
books with identifier & 0 & 0 & 0 & 0 \\ \hline
journals verified open access & 1 & 0.07 & 0 & 0.07 \\ \hline
Google Books with full/public domain access & 1 & 0 & 1 & 0 \\ \hline
Google Books with partial access & 0.5 & 0 & 0 & 0.2 \\ \hline \hline
\textbf{Article Score} & & 2.07 & 3 & 2.27 \\ \hline
\end{tabular}
\label{tab:table1}
\end{table}

Table \ref{tab:table2} reports the weighting scheme for each model. For Model 2, we weighted technical measures of verifiability more heavily. Model 3 instead weighted practical elements more heavily. Finally, Model 4 used baseline weighting for technical and practical elements, and added the two identifier categories, to reward the presence of electronic identification numbers. For each model the weighting changed the possible score for each article. For example, ``Bugatti'' scored a 3 in the baseline model, 5 in model 2, 4 in model 3, and 3.33 in model 4. 

\begin{table}[ht]
\caption{
{\bf Weighted components for each model.}}
\begin{tabular}{r c c c c}
\textbf{Proportion of} & \textbf{Model 1} & \textbf{Model 2} & \textbf{Model 3} & \textbf{Model 4}\\
ISBNs valid & 1 & 2 & 1 & 1 \\ \hline
Google Books links valid & 1 & 2 & 1 & 1 \\ \hline
journals with DOI & 0 & 0 & 0 & 1 \\ \hline
books with identifier & 0 & 0 & 0 & 1 \\ \hline
journals verified open access & 1 & 1 & 2 & 1 \\ \hline
Google Books with full/public domain access & 1 & 1 & 2 & 1 \\ \hline
Google Books with partial access & 0.5 & 0.5 & 1 & 0.5\\ \hline
\end{tabular}
\label{tab:table2}
\end{table}

Article scores are only directly comparable within a model, so we ranked articles according to their individual scores under each model to get a sense of inter-model consistency, and then compared rank across models. This can be visualized as a scatter plot, with the x-axis representing articles 1 to 5,000 in descending order of score according to Model 1. Each article's corresponding rank in the model being compared is then plotted on the y-axis. 

\begin{figure}[h]
       \includegraphics[width=\columnwidth]{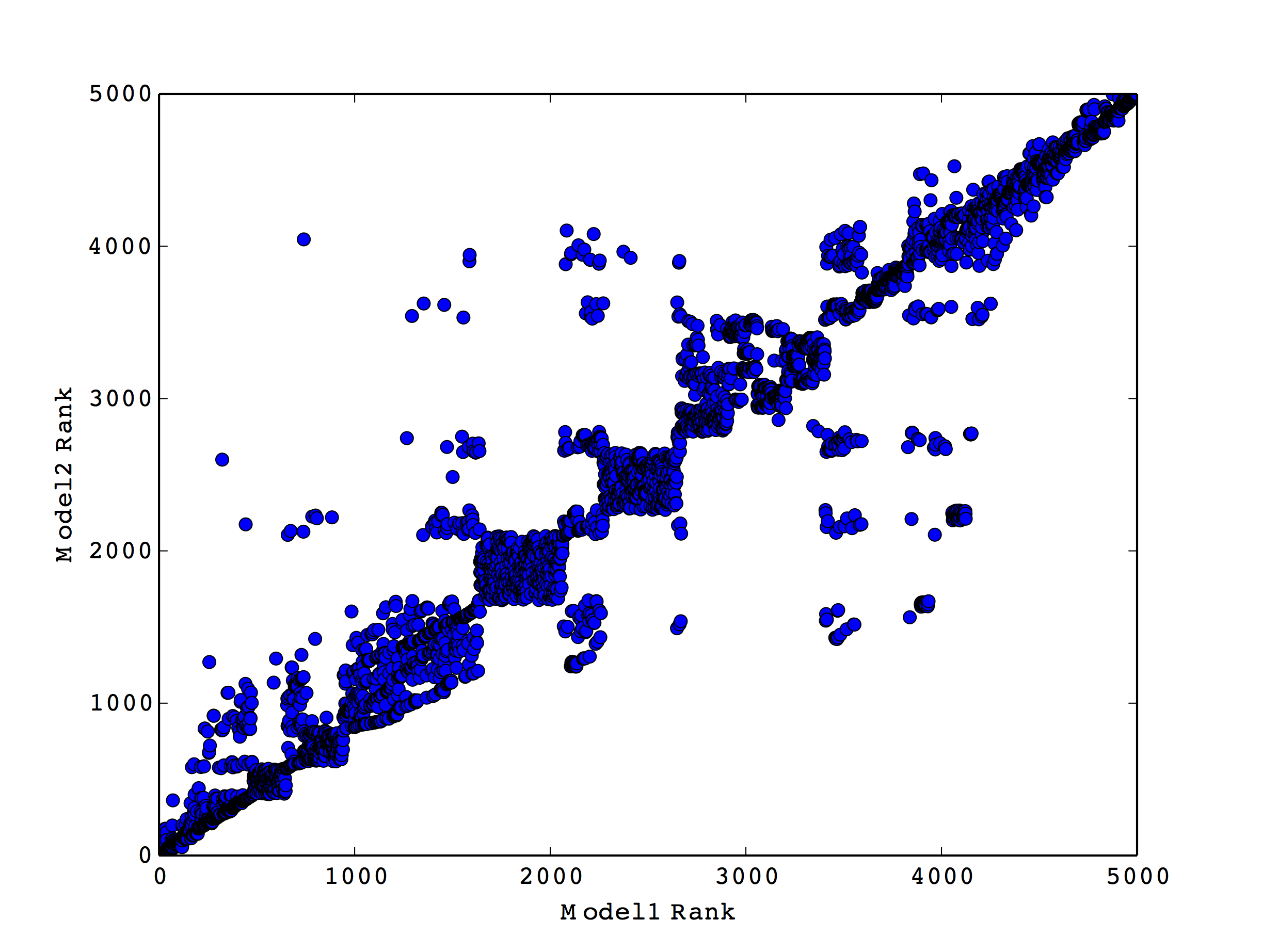}
  \caption{\bf Change in article verifiability rank, baseline model (Model 1) vs.\ Model 2.}~\label{fig:figure4}
\end{figure}

\begin{figure}[h]
      \includegraphics[width=\columnwidth]{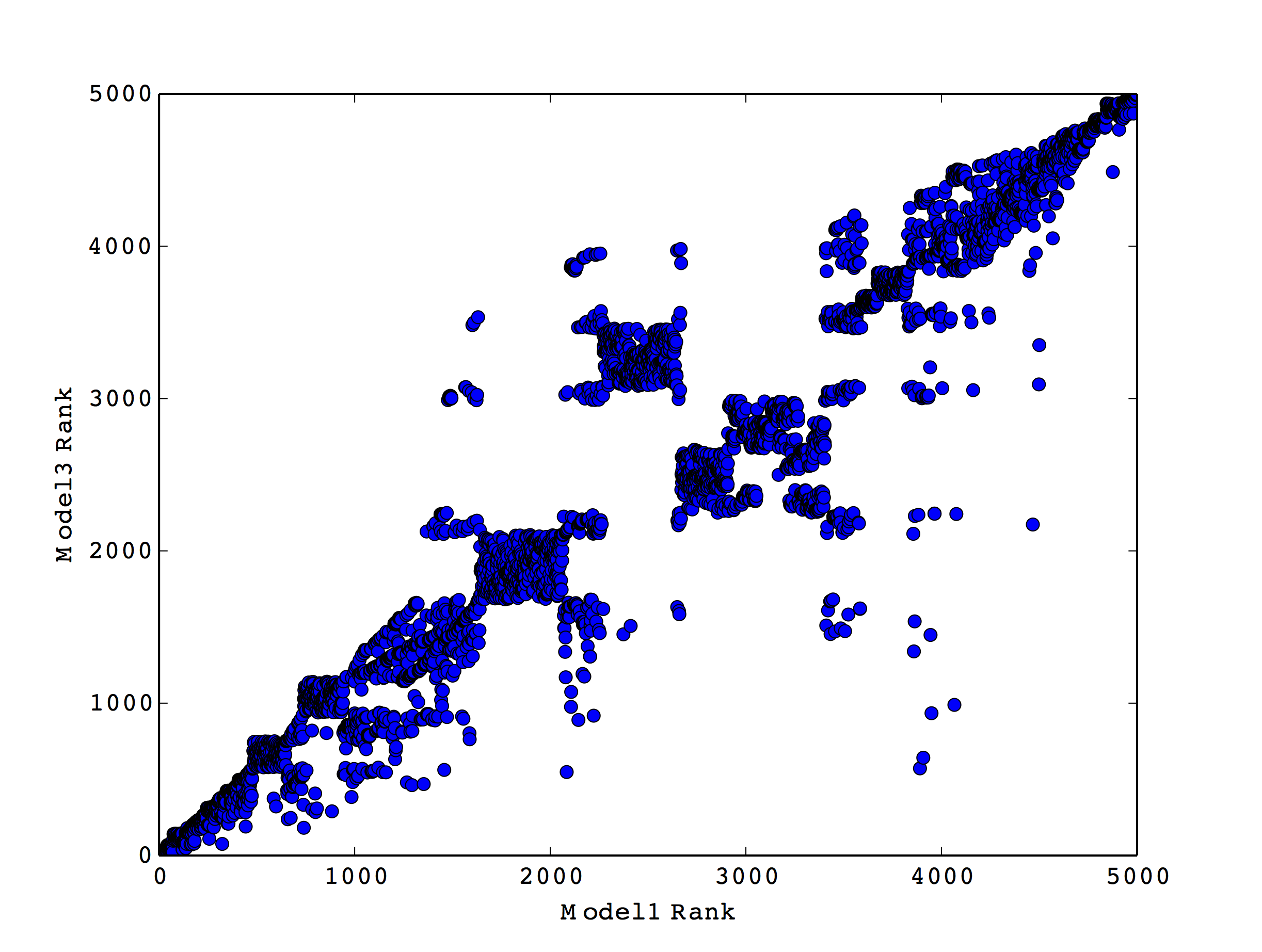}
  \caption{\bf Change in article verifiability rank, baseline model (Model 1) vs.\ Model 3.}~\label{fig:figure5}
\end{figure}

\begin{figure}[h]
  \centering
      \includegraphics[width=\columnwidth]{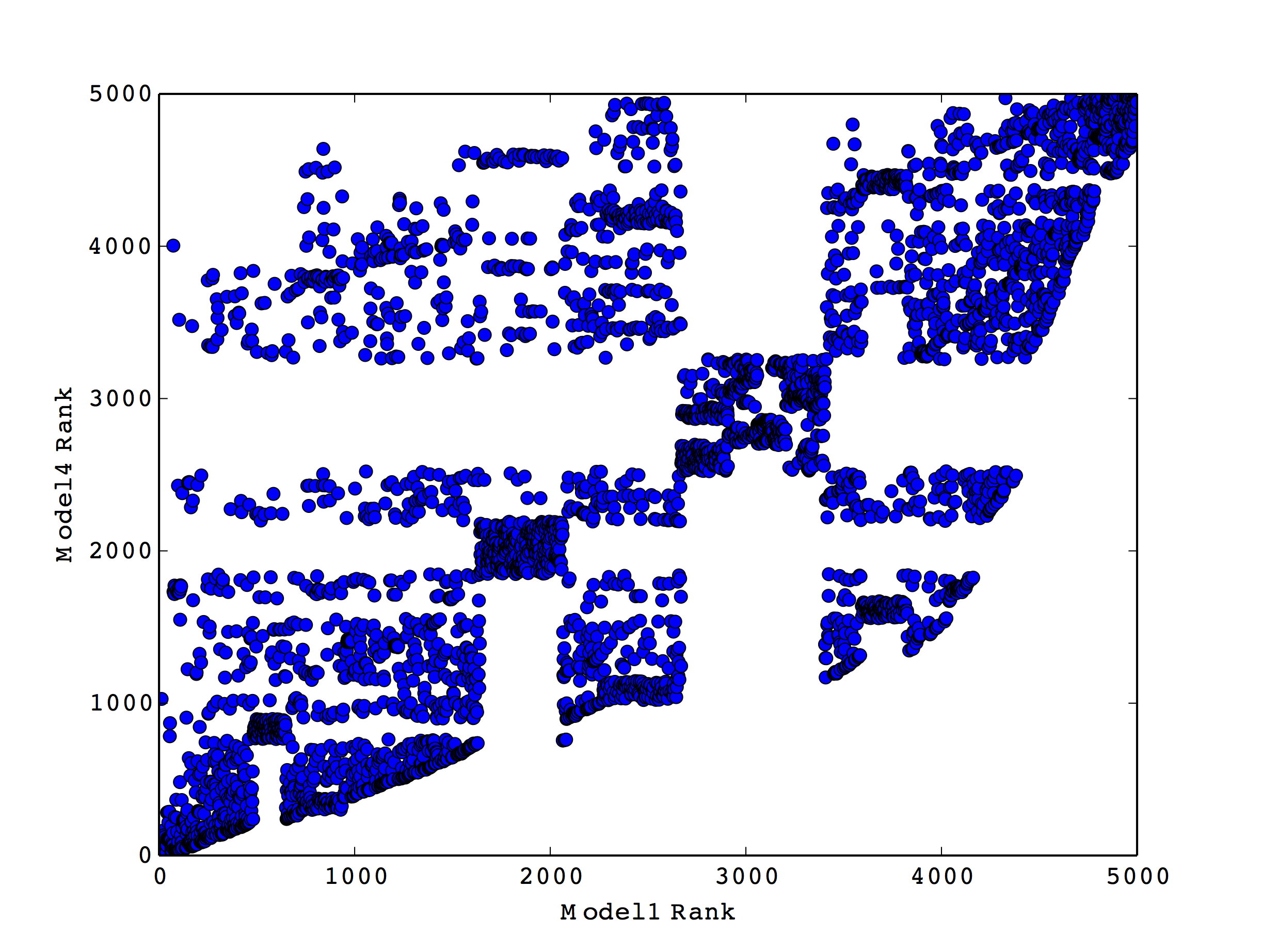}
  \caption{\bf Change in article verifiability rank, baseline model (Model 1) vs.\ Model 4.}~\label{fig:figure6}
\end{figure}

As Figs.~\ref{fig:figure4} and \ref{fig:figure5} illustrate, Models 2 and 3 show relative consistency in ranking with Model 1. By contrast, as Fig.~\ref{fig:figure6} illustrates, Model 4 (with added identifier rankings) shows some significant variability in ranking. Block-like structures in the plot arise from regions of uniform scoring according to Model 1.

To get a sense of the factors underlying divergences in ranking between models, some specific examples are illustrative. The largest gain in rank from Model 1 to Model 2 was the article ``Arbitration,'' which gained 2,294 spots, from having a score ranked 3,931 to a score ranked 1,637. This gain makes sense in light of Model 2's emphasis on citation validity, as both of this article's ISBNs and both of its Google Books IDs were valid. The greatest loss in rank was by the article ``Microwave,'' which dropped 3,305 spots from rank 741 to 4,046. One of its two ISBNs was invalid, and one of its three Google Books links was broken. 

Comparing Model 1 and Model 3, the greatest gain in article rank was a 3,318 spot jump by ``Glycerol'' from rank 3,891 to rank 573. This article's only ISBN was invalid, explaining a low ranking under Model 1, but its one Google Books ID was fully viewable, raising the article's relative score under Model 3's emphasis on practical verifiability. The greatest drop in this comparison was the article ``Nero,'' which dropped 1,903 places from rank 1,632 to 3,535, hurt under greater emphasis on practical verifiability with three out of its five Google Books IDs being completely unavailable for free online viewing. 

Comparing Model 1 and Model 4, the greatest gain in rank was by ``Pneumothorax,'' which jumped 2,497 places from rank 3,856 to rank 1,359. With Model 4's added weighting for the presence of identifiers, this article was helped by the fact that all 24 of its journals had electronic identification (DOI, or confirmed open access), and seven out of its nine book links contained either a valid ISBN or Google Books ID. The greatest drop in rank under Model 4 was by ``Bugatti,'' which dropped 3,931 places from rank 74 to rank 4,005. Both of its journal citations had no electronic identification and two out of its three books contained neither an ISBN nor Google Books link.

\section*{Discussion}
We have presented an approach to measuring verifiability that illuminates potential problems with technical and practical verifiability alike. From the perspective of overall quality of references in Wikipedia, these findings might seem encouraging. Almost 85\% of book references and 87\% of journal references contain valid, standardized identifiers such as ISBN, Google Books ID, DOI, or direct links to known open access journals. However, many of these references are not practically verifiable, with only 12.8\% of journal citations confirmed as open access in arXiv or PMC, and only 12.4\% of Google Books links confirmed as fully viewable.

Four caveats are worth noting that suggest interesting directions for future research. First, we did not check to see that these standardized identifiers correctly matched the provided textual reference information. Second, these findings are for the 5,000 English-language articles that draw the most attention, so the overall quality presented here might differ from the modal Wikipedia article, or from articles in other languages. Third, the models only consider obviously Open Access sources such as PubMed and arXiv, and might productively be expanded to include other sources known to be Open Access (e.g. listed in the Directory of Open Access Journals). Fourth, the measures in these simple models do not account for the small number of references in Wikipedia articles, and might productively be extended in future research to be more robust.


Future research on Wikipedia might also examine in more detail the variation in verifiability scores between individual articles in different models. Using fractions makes the models more robust for articles with many references, so rankings for a single article with few book or journal references can change significantly even if the change in number or validity of references is small in absolute terms. This suggests future opportunities for considering reference density and reference quality together in the study of verifiability. One possible direction would be investigating effects of genre or category on verifiability. There may well be informal, genre-specific editorial expectations that favor one model of verifiability over another. Similarly, comparison of article verifiability rankings against Wikipedia's internal article quality rankings could provide useful insight. While previous work has noted a relationship between article age and density of references~\cite{Chen:2012cl}, a consideration of reference quality might illuminate more complex relationships between article quality and article maturity. 

A more practical direction would be to incorporate an article-level verifiability metric into the Wikipedia browsing experience, allowing users to compare the empirical reality of verifiability against broader policy expectations. Connecting verifiability to user experience would also address verifiability as a potential source of user inequality and bias in Wikipedia articles. The burden of satisfying the verifiability metric currently falls on editors who may have very different access to, and preferences for, reliable knowledge~\cite{Wikipedia:2015ij}. Making verifiability visible to users could encourage wider participation by users with different perspectives on access to knowledge.

Our approach to constructing a flexible and customizable verifiability metric helps make visible potential problems of verifiability, and increases the possibility for improving verifiability for all users in Wikipedia. But what is possible for Wikipedia is also possible for many other online information sources. For example, this approach could be extended to measure the practical verifiability of scientific papers by looking at whether their supporting citations and data are readily available for review. Similarly, this approach could be extended to a browser extension that scrapes any citation or reference on a web page and calculates that page's technical and practical verifiability. But however this approach is extended in the future, measuring verifiability will help address variations in the quality of references in online information and, ideally, improve their overall quality.

\section*{Acknowledgments}
We gratefully acknowledge the support of the Neukom Institute for Computational Science. We also acknowledge the helpful feedback received from several anonymous readers.


%
%
%
%

\bibliography{WV1.2}

\pagebreak

\newpage

\end{document}